\documentclass[11pt]{article}
\usepackage{amssymb}
\usepackage{amsmath, amsthm}
\newcommand{\be}{\begin{equation}}
\newcommand{\ee}{\end{equation}}
\begin{document}
\def\theequation{\arabic{section}.\arabic{equation}}
\begin{titlepage}
\title{Palatini $f(R)$ gravity as a fixed point}
\author{Valerio Faraoni \\ \\
{\small \it Physics Department, Bishop's University}\\
{\small \it 2600 College St., Sherbrooke, Qu\'{e}bec, Canada 
J1M~1Z7}\\
{\small \it email~~ vfaraoni@ubishops.ca}
}
\date{} \maketitle
\thispagestyle{empty}
\vspace*{1truecm}
\begin{abstract}
In the context of modified gravity, we point out how 
the Palatini version of these theories is singled out as a   
very special case corresponding to the unique fixed 
point of a  transformation involving a special conformal 
rescaling of the metric. This mathematical peculiarity signals 
deeply rooted problems which make the theory unphysical.
\end{abstract}
\begin{center}
\end{center} 
\begin{center} Keywords: modified gravity, Palatini formalism. 
\end{center} 
\end{titlepage} 
\clearpage 
\setcounter{page}{2} 
\section{Introduction}
\setcounter{equation}{0}

Among the multitude of efforts devoted to explaining and 
modelling the 
current acceleration of the cosmic expansion discovered with 
type Ia supernovae \cite{SN}, modified (or $f(R)$) gravity has 
received 
much attention.  This class of theories aims at disposing of the 
concept of dark energy by assuming that, instead, we may be 
observing the first deviations from Einstein's general 
relativity on cosmological scales (\cite{modifiedgravity}, see 
\cite{review} for a review).  
Modified gravity is described by the action
\be \label{metricaction}
S=\frac{1}{2\kappa} \int d^4 x \sqrt{-g} \, f(R) +S^{(m)} 
\left[ 
g_{ab}, \psi \right] \;,
\ee
where $ S^{(m)}=\int d^4x \, \sqrt{-g} {\cal L}^{(m)}$ is the 
matter 
action and $\psi$ collectively denotes 
the matter fields, $g$ is the determinant of the 
spacetime metric $g_{ab}$, $\kappa\equiv 8\pi G$, where $G$ is 
Newton's constant (we follow the notations of \cite{Wald} and 
use units in which $G=c=1$), and 
$f(R)$ is a (generically nonlinear and 
twice differentiable) function of the Ricci curvature $R$ which 
generalizes the Einstein-Hilbert action, to which 
it reduces when 
$f(R)=R$.  Modified gravity comes in three versions: the {\em 
metric formalism} in which the connection is the metric 
connection of $g_{ab}$; the {\em Palatini formalism} 
\cite{Vollick} in which 
the metric and the connection are independent variables ({\em 
i.e.}, the connection $\Gamma^a_{bc}$ is not the metric 
connection of $g_{ab}$), the Ricci tensor ${\cal R}_{ab} $ is 
built out of this connection, and ${\cal R}\equiv g^{ab}{\cal 
R}_{ab}$ is the Ricci curvature appearing in the action. The 
latter should properly be written as 
\be \label{Palatiniaction}
S=\frac{1}{2\kappa} \int d^4 x \sqrt{-g} \, f({\cal R}) 
+S^{(m)} 
\left[ g_{ab}, \psi \right] \;.
\ee
In the Palatini formalism, the matter action $S^{(m)}$ is 
independent of the (non-metric) connection $\Gamma^a_{bc}$; a 
third version 
of $f(R)$ gravity, the {\em metric-affine formalism}, allows 
$S^{(m)}$ to depend explicitly on this connection. This version 
is 
little studied \cite{metricaffine} and will not be considered 
here. We focus on metric and Palatini $f(R)$ gravity, which give 
rise to fourth order (in the metric) and second order field 
equations, respectively. The field equations of metric $f(R)$  
gravity are
\be \label{metf}
 f'(R)R_{ab}-\frac{1}{2}f(R)g_{ab}- 
\left( \nabla_a \nabla_b -g_{ab} \Box\right) f'(R)= 
\kappa \,T_{ab} \;, 
\ee
where a prime denotes differentiation with respect to $R$ and  
$ T_{ab}=\frac{-2}{\sqrt{-g}}\, \frac{\delta
S^{(m)} }{\delta g^{ab} } $. The Palatini field equations are 
\be
f'({\cal R}) {\cal R}_{ab}-\frac{1}{2}f({\cal 
R})g_{ab}=\kappa \, T_{ab} \;, \;\;\;\;\;\; 
\bar\nabla_\lambda\left[ \sqrt{-g}f'({\cal 
R})g^{ab}\right]=0 \;,
\ee
where $\bar{\nabla}_c $ denotes the covariant 
derivative operator of the independent connection  
$\Gamma^a_{bc}$.

Recently, it has been pointed out that Palatini $f(R)$ gravity 
is not viable because of two serious shortcomings:\\ 
a)~when trying to build  a stellar model in the weak-field limit 
using very reasonable (polytropic) fluids, a singularity  
in the curvature invariants appears at the star's surface, which 
is related to the impossibility of matching interior and 
exterior (vacuum) solutions.  This feature has been traced back 
to the fact that the metric depends on derivatives of order 
higher than first of the matter fields 
entering the field equations. As a result, discontinuities in 
the matter distribution are not smoothed out by an 
integral, as  in conventional theories, but the metric  
depends on the matter fields and their derivatives, causing 
singularities in the curvature that are physically 
unacceptable~\cite{BarausseSotiriouMiller}.\\
b) The initial value problem is not well-formulated nor 
well-posed for Palatini $f(R)$ gravity  
\cite{LanahanFaraoni}. Nevertheless, many papers still 
appear 
on Palatini $f(R)$ gravity, and here we try to understand its 
very special features from a completely different perspective.

It is well-known that, when $f''(R)\neq 0$, metric $f(R)$ 
gravity is dynamically equivalent to an $\omega=0$ Brans-Dicke  
(hereafter BD) theory \cite{BD} for the massive scalar degree of 
freedom 
$\phi \equiv 
f'(R)$, while Palatini $f(R)$ gravity is equivalent to an 
$\omega=-3/2$ BD theory \cite{STequivalence}. In the 
metric formalism, by introducing an auxiliary field $\chi$, one 
can consider the action (dynamically equivalent 
to~(\ref{metricaction}) if $f''\neq 0$)
\be
S_0=\frac{1}{2\kappa} \int d^4 x \sqrt{-g} \left[ f(\chi) 
+f'(\chi) \left( R-\chi \right) \right]
+S^{(m)} \left[ 
g_{ab}, \psi \right] \;,
\ee
the variation of which with respect to $\chi$ yields 
$f''(R)\left( \chi - R \right)=0$. By defining the scalar $\phi 
\equiv f'(\chi)$, this action becomes
\be \label{8}
S_0 =\frac{1}{2\kappa} \int d^4 x \sqrt{-g} \left[ \phi R 
-V(\phi) 
\right] +S^{(m)} \;,
\ee
where $
V(\phi)=\chi(\phi)\phi -f\left( \chi(\phi) \right) $. This is a 
BD action with Brans-Dicke parameter 
$\omega=0$ and potential $V$. Similarly, in the Palatini 
formalism, the 
action~(\ref{Palatiniaction}) becomes
\be
S_{Palatini}=\frac{1}{2\kappa} \int d^4 x \sqrt{-g} \left[ 
\phi {\cal R} - V( \phi) \right]
+S^{(m)} \left[ 
g_{ab}, \psi \right] \;,
\ee
but now ${\cal R}$ is not the Ricci curvature $R$ of the metric 
connection. The relation between the two is
\be
\label{confrel2}
{\cal R}=R+\frac{3}{2\left[ f'({\cal 
R})\right]^2}\left[\nabla_a f'({\cal 
R})\right]\left[ \nabla^a f'({\cal 
R})\right]+\frac{3}{f'({\cal R})} \Box f'({\cal R}) \;,
\ee
from which one obtains, apart from irrelevant boundary terms,
\be \label{12}
S_{Palatini}=\frac{1}{2\kappa} \int d^4 x \sqrt{-g} \left[ 
\phi  R +\frac{3}{2\phi}\, \nabla^c \phi\nabla_c\phi - V( \phi) 
\right] +S^{(m)}  \;,
\ee
an $\omega=-3/2 $ BD theory, which is seldom considered in 
the literature \cite{ST-3/2}.  Note that the Ricci scalar $R$ 
appearing in eq.~(\ref{12}) is constructed with the Ricci tensor 
$R_{ab}$ of the metric connection of $g_{ab}$, and 
differs from the scalar ${\cal R}$ used earlier.

\section{Palatini $f(R)$ gravity as a fixed point}
\setcounter{equation}{0}

The general form  of the BD action in the Jordan frame 
is
\be \label{BDaction} 
S_{BD}=\frac{1}{2\kappa}\int d^4x \sqrt{-g}
\left[ \phi R - \frac{\omega}{\phi} \, g^{ab} 
\nabla_{a}\phi 
\nabla_{b}\phi -V(\phi)  \right] + S^{(m)} \; , 
\ee 
and the corresponding field equations are 
\be  
R_{ab}-\frac{1}{2} g_{ab}
R=\frac{\kappa}{\phi} \, T_{ab} + \frac{\omega}{\phi^2} 
\left(
\nabla_{a}\phi \nabla_b \phi -\frac{1}{2} g_{ab}
\nabla^{c}\phi \nabla_c\phi \right) +\frac{1}{\phi} 
\left(
\nabla_a \nabla_b \phi-g_{ab} \Box \phi \right) 
-\frac{V}{2\phi}\, g_{ab}  \; , 
\ee 

\be \label{Box}
\left(3+2\omega \right) \Box \phi =\kappa T^{(m)} +\phi \, 
\frac{dV}{d\phi} -2V \; ,
\ee 
where $T$ is the trace of the matter stress-energy tensor. 
Here we use the equivalence between $f(R)$ and BD 
gravities and an invariance property of the latter to elucidate 
the very special role played by Palatini modified gravity in the 
broader spectrum of $f(R)$ and BD theories. Let us consider the  
gravitational 
sector of the theory: under the conformal 
transformation\footnote{The  transformation~(\ref{trans1}) and 
(\ref{trans2}) was used in \cite{Faraoni} to  study the 
$\omega\rightarrow \infty$ limit 
of BD theory to general relativity, which may fail in the 
presence of conformally invariant matter.} 
\be \label{trans1}
 g_{ab} \longrightarrow
\tilde{g}_{ab}=\Omega^2 g_{ab} \; , \;\;\;\;\;\;
 \Omega=\phi^{\alpha}  \;\;\;\;\;\;\;\;(\alpha \neq 1/2), 
\ee 
and the scalar field redefinition
\be \label{trans2}
\phi \longrightarrow \sigma= \phi^{1-2\alpha} \;,
\ee 
and using the transformation property of the Ricci scalar under 
conformal transformations \cite{Wald}
\be
 \tilde{R}=\Omega^{-2} \left( R+\frac{6\Box \Omega}{\Omega}
\right) \;, 
\ee 
the BD action is rewritten as 
\be
S_{BD}=\frac{1}{2\kappa}\int d^4x \sqrt{-\tilde{g}}
\left[ \sigma \tilde{R} -\frac{\tilde{\omega}}{\sigma}\,  
 \tilde{g}^{ab} 
\tilde{\nabla}_a\sigma 
\tilde{\nabla}_b \sigma -U(\sigma) \right] + S^{(m)} \; , 
\ee 
where
\be \label{newomega}
 \tilde{\omega}=\frac{\omega -6\alpha \left( \alpha -1
\right)}{\left( 1-2\alpha \right)^2} \; , \;\;\;\;\;\;\;\;\;
U(\sigma)=V\left( \sigma^{\frac{1}{1-2\alpha}} \right) \;, 
\ee
{\em i.e.}, the gravitational
part of the BD action is invariant in form under the 
transformation (\ref{trans1}) and (\ref{trans2}). This 
restricted conformal 
invariance property is well-known an has been  likened 
to  the conformal invariance of string theories at high energies 
\cite{Turner, mybook} (remember that the 
low-energy limit of the bosonic string theory is an $\omega=-1$  
BD theory \cite{bosonicstring}). The value $-3/2$ of the 
parameter $\omega$ is special; in fact,
the function $\tilde{\omega}\left( \omega, \alpha \right)$ is 
singular at $\alpha=1/2$ when $\omega \neq -3/2$, has two 
branches for $\alpha > 1/2 $ and $\alpha < 1/2$, and  
\be
\lim_{\alpha \rightarrow 1/2} \tilde{\omega}= 
\left\{
\begin{array}{lc}
& +\infty  \;\;\;\; \;\;{\mbox if}\;\;\; \omega>-3/2 \;, \\
& -\infty  \;\;\;\; \;\;{\mbox if}\;\;\; \omega<-3/2 \;, \\
& -3/2  \;\;\;\; \;\;{\mbox if}\;\;\; \omega=-3/2 \;, 
\end{array} \right. 
\ee
It is  $\tilde{\omega}<0$ if 
$\omega<-3/2$, $\tilde{\omega}\geq 0$ if $\omega>-3/2$ and 
$\alpha_1 \leq \alpha \leq \alpha_2$, and $\tilde{\omega} < 0$ 
if $\omega >-3/2$ and 
$\alpha< \alpha_1 $ or $\alpha >\alpha_2$, where $
\alpha_{1,2} =\frac{1}{2} \left( 1\pm \sqrt{ 1+\frac{3\omega}{2} 
}\,  \right) $. It is interesting to look for fixed points of 
the 
transformation~(\ref{trans1}) and (\ref{trans2}) as one moves 
in the space of BD theories $\left( g_{ab}^{( \omega )}, 
\phi^{( \omega )}, V(\phi) \right)$;  the 
theory is  already invariant in form under this transformation, 
and we define as a {\em fixed 
point} a BD theory identified by the condition 
that the BD parameter does not change, $\tilde{\omega}=\omega$. 
The potential $U(\sigma)$ will, in general, have a different 
functional form from the potential $V(\phi)$, but it seems 
reasonable to 
allow for this because the arbitrariness in the choice of the 
function $f(R)$ (or $f({\cal R} )$) implies arbitrariness in the 
choice of the potential.

Apart from the trivial cases in which the 
transformation~(\ref{trans1}), (\ref{trans2}) 
reduces to the identity (corresponding to $\alpha=0$ or 
$\alpha=1$), this equality is satisfied for $\omega=-3/2$. 
Palatini $f(R)$ gravity, corresponding to an $\omega=-3/2$ BD 
theory, is therefore singled out as the unique  fixed 
point of the transformation~(\ref{trans1}), (\ref{trans2}). It 
is not  difficult to understand why this case is so special: the 
dynamical field  equation~(\ref{Box}) for the BD scalar $\phi$ 
degenerates 
into the  algebraic identity $2V-\phi V'= \kappa \, T$ for this 
value of $\omega$. The dynamical equation for $\phi$ 
disappears, leaving this field with no dynamical 
role.\footnote{There is an exception: the case in which $\Box 
\phi=0$, which includes general relativity (for $\phi=$const.) 
and harmonic $\phi$-waves.}  It is 
exactly this fact that makes the Cauchy problem for Palatini 
$f(R)$ gravity ill-formulated and, therefore, ill-posed: because 
there is no expression for $\Box\phi$ that can be substituted by 
the matter trace $T$ in the $3+1$ decomposition of the field 
equations, second derivatives of $\phi$ can not be eliminated 
from these equations, contrary to the case  $\omega\neq -3/2$ 
 \cite{LanahanFaraoni}. The fact that 
Palatini modified gravity has this (restricted) conformal 
invariance property singles it out among modified gravity 
theories, and nowhere is this more evident than in the 
equivalent BD theory.

An equivalent way of looking at this issue is the consideration 
of  the Einstein frame formulation of this 
theory. It is well known that, by performing a conformal 
transformation of the metric 
and a scalar field redefinition (different from~((\ref{trans1}) 
and~(\ref{trans2})), and given  instead by
\begin{eqnarray}
&& g_{ab} \rightarrow \tilde{g}_{ab}=\phi \, g_{ab} 
\;\;\;\;\;\;\;\;\; (\Omega=\sqrt{\phi} \,) \;, \\
&&\nonumber \\
&&\phi \rightarrow \tilde{\phi}=\int  \left| 3+2\omega
\right|^{1/2}  \, \frac{d\phi}{\phi}  \;,
\end{eqnarray}
a BD theory is mapped into its Einstein frame representation 
$\left( \tilde{g}_{ab}, \tilde{\phi} \right)$ in which the 
action~(\ref{BDaction}) assumes the form
\be \label{Einsteinaction} 
S_{BD}=\frac{1}{2\kappa} \int d^4  \sqrt{-\tilde{g}} \left[ 
\tilde{R}  -\frac{1}{2} \,
\tilde{g}^{ab} \tilde{\nabla}_a \tilde{\phi}
\tilde{\nabla}_b \tilde{\phi} -\frac{ 
V(\phi(\tilde{\phi}))}{\phi^2}  +{\cal L}^{(m)} \left[ 
\phi^{-1} 
g_{ab}, \psi \right] \right] \; ,  
\ee 
which corresponds to general relativity with a 
scalar $\tilde{\phi}$  minimally coupled to the 
curvature, but which exhibits an ``anomalous'' coupling of 
to the matter fields\footnote{Usually, only values of 
the BD parameter $\omega>-3/2$ are considered in the literature 
and this transformation appears without the absolute 
value in the argument of  the square root.} 
$\psi$ (unless the latter  are conformally invariant). The 
Jordan and Einstein frames are 
physically equivalent representations of the same theory 
\cite{Dicke, Flanagan} (at least at the classical level 
\cite{FaraoniNadeau}). The definition of the Einstein frame 
scalar $\tilde{\phi}$ breaks down as $\omega \rightarrow 
-3/2$. One can still perform the conformal transformation of the 
metric without redefining the field $\phi$, thus obtaining the 
Einstein frame action
\be  
S_{Palatini}=\frac{1}{2\kappa} \int d^4  \sqrt{-\tilde{g}} 
\left[  \tilde{R}   -\frac{ 
V( \phi ) }{\phi^2}  +\frac{ {\cal L}^{(m)} \left[ 
\phi^{-1} g_{ab}, \psi \right] }{  \phi^2 } \right] \; ,
\ee 
the variation of which with respect to $\phi$ leads again to 
$2V-\phi V'=\kappa T$. In other words, there is no dynamical 
equation for the scalar field $\phi$, which is appended to 
the gravitational and matter  actions and can be assigned 
arbitrarily {\em a priori}. 
The scalar $\phi$ becomes unphysical and, therefore, this 
theory is unattractive from 
the physical point of view.

In some sense, Palatini $f(R)$ gravity appears to be 
unphysical because it corresponds to forcing conformal 
invariance onto a field that is intrinsically 
non-dynamical and whose existence is not justified  
from the physical point of view. $\phi$ is the inverse of the  
effective gravitational coupling; in the 
transformation~(\ref{trans1}), the latter
acts as a scaling factor for all lengths, times, and masses.  
To preserve the {\em form} of the field equations, this 
effective coupling itself needs to be changed  
appropriately (eq.~(\ref{trans2})). In the absence of 
matter and of the potential (or with a quadratic potential), 
the dynamics of BD theory are left unchanged by these 
rescalings, at the price of changing the {\em value} of the BD 
parameter (eq.~(\ref{newomega}). This parameter, which 
weights the relative importance of the kinetic and the $\phi R$ 
terms in the BD action, can be changed by a large amount. Now, 
requiring that also the {\em value} of $\omega$ be left 
invariant by the transformation~(\ref{trans1}),  
(\ref{trans2}) is simply too much: the only way to achieve 
this is by losing completely the dynamics of $\phi$, which can 
then be assigned  arbitrarily in infinitely many 
different ways (for example, two prescriptions  for $\phi$ 
and its derivatives could coincide on an initial hypersurface 
$\Sigma$  and differ in the future  domain of 
dependence of $\Sigma$, resulting in the non-uniqueness of 
solutions that makes the Cauchy problem ill-posed 
\cite{LanahanFaraoni}). Downgraded to an  
auxiliary field, the  effective gravitational coupling 
$\phi^{-1}$ becomes completely arbitrary, which defeats the 
original purpose of its  
introduction  ({\em i.e.}, having it determined by 
the distribution of all masses in the cosmos) \cite{BD}, 
and destroys also the more modern motivation of scalar-tensor 
gravity with $\phi$ entering as a 
dynamical dilaton. $\phi$ is then assigned in a manner external 
to the theory and can not be determined internally. In this 
situation, the only meaningful choice is $\phi=$const., which 
reproduces general relativity.

It is instructive to examine a toy model in particle 
dynamics which mimics Palatini $f(R)$ gravity and was proposed  
in \cite{Kaloper}. This is given by the action
\be \label{toy}
S\int dt \left( \frac{wx \dot{Q}^2}{2Q} 
-xQ\dot{Y}-\frac{x}{2}\, QY^2 -xJ \right) \;,
\ee
dependent on the parameter $w$, where $x$ corresponds to 
$\sqrt{-g}g^{ab}$, $Y$ to the 
connection $\Gamma$, and $Q$ to $\phi$. The analog of the 
transformation to the Einstein frame is the change of variables 
$z \equiv xQ$, $ Q\equiv \mbox{e}^{q/\sqrt{w}}$, which casts 
the action~(\ref{toy})  into \cite{Kaloper}
\be \label{EFtoy}
S= \int dt {\cal L}= \int dt \left( \frac{z}{2}\,  
\dot{q}^2  
-z\dot{Y}-\frac{z}{2}\, Y^2 -z \, \mbox{e}^{-q/\sqrt{w}} \, J 
\right) \;.
\ee
(A proper treatment should include a Lagrange 
multiplier: we do not show it here and refer the 
reader to the discussion of \cite{Kaloper}.)

What is, in this toy model,  the analog of our
transformation~(\ref{trans1}) and (\ref{trans2})? Ignoring 
again matter (represented by $J$) for simplicity, it is 
straightforward to check that this is given by
\be\label{toytrans}
z=xQ^{\alpha} \;, \;\;\;\;\;\;\;\;
Q=q^{\frac{1}{\alpha-1} } \;, \;\;\;\;\;\;\;\;
p=1/q \;,
\ee
which transforms the Lagrangian density  into 
\be
{\cal L}=\frac{\tilde{w} \, \dot{p}^2 }{2p} 
-zp\dot{Y}-\frac{z}{2} \, p \, Y^2 \;, \;\;\;\;\;\; 
\tilde{w}=\frac{w}{\left( \alpha -1 \right)^2} \;, 
\ee
{\em i.e.}, the action~(\ref{toy}) is invariant {\em in form} 
under 
the transformation~(\ref{toytrans}). Excluding the 
trivial  situation  $\alpha=2$ and $Q=q$, there is only one 
occurrence in which also the {\em value} of the parameter $w$ 
is left unchanged, namely the case $w=0$. In this case, the 
dynamics of the variable $Q$  are lost. This situation appears 
rather trivial in the toy model employed, but it shows that 
imposing too strict of a requirement (the invariance in  
value of the parameter $w$ mimicking $\omega$) leads to an 
unphysical situation. Our main point, that the very special 
role played by the conformally invariant value $-3/2$ of the 
BD parameter $\omega$ leads to an unphysical theory, is 
exemplified by the unphysical zero value  of the parameter $w$. 
The very special role of these parameter values from the 
mathematical point of view corresponds to unphysical 
situations.

To complete the analogy between $\omega=-3/2$ BD theory and 
the toy model, one notes that the analog of the transformation 
to the ``Einstein frame'' is spoiled when $w=0$. The variable 
$ Q\equiv \mbox{e}^{q/\sqrt{w}}$ can not be defined, but the 
other variable $z \equiv xQ$ is still well defined and one can 
define an ``Einstein frame'' physically equivalent to the 
``Jordan frame'' in  terms of $z$ and $Q$. The ``Einstein 
frame'' action is (apart from the Lagrange multiplier discussed 
in \cite{Kaloper})
\be
S=\int dt \left[ -z \left( \dot{Y} + \frac{Y^2}{2} 
\right)+Q^{-1}J 
\right] \;,
\ee
where the last term exhibits the non-minimal coupling of the 
non-dynamical field $Q$ to matter.

We conclude that the very special value $-3/2$ of the BD 
parameter from the mathematical point of view signals the 
unphysical features of this theory and of Palatini $f(R)$ 
gravity. A possible cure (cf., {\em e.g.}, 
\cite{BarausseSotiriouMiller}) would be to generalize 
the $f(R)$  action by including terms in $R_{ab}R^{ab}$, 
$R_{abcd}R^{abcd}$, or other invariants of 
the Riemann tensor. 
This would have the effect of raising the order of the field 
equations by two and restore non-trivial dynamics. However, 
unless the extra terms appear in the Gauss-Bonnet combination, 
one will be faced with the well known Ostrogradski instability.

\section*{Acknowledgments}

The author is grateful to Thomas Sotiriou for a helpful 
discussion, to a referee for useful comments,  and to the 
Natural Sciences and Engineering Research 
Council of Canada (NSERC) for financial support.

\vskip1truecm

\end{document}